\documentclass{eas}
\usepackage{graphicx}
\begin{document}

\title{SMC3 as a test to the binary evolution} \thanks{KI gratefully acknowledgements support from the LOC. This study has been supported in part by the Polish MNiSW grant 0136/DIA/2014/43.}
\author{Krystian I{\l}kiewicz}\address{Nicolaus Copernicus Astronomical Centre, Bartycka 18, 00716 Warsaw, Poland}
\author{Joanna Miko{\l}ajewska}\sameaddress{1}
\author{Krzysztof Belczy{\'n}ski}\address{Warsaw University Observatory, Al. Ujazdowskie 4, 00-478 Warszawa, Poland}
\begin{abstract}
SMC 3 is one of the most interesting symbiotic stars. This binary contains a bright K-type giant transferring mass to a massive white dwarf comanion, which makes it is a very promising SN Ia candidate. We discuss the evolutionary status of the system using results of population synthesis code.
\end{abstract}
\maketitle
\section{Introduction}
SMC 3 is one of the symbiotic systems in the Magellanic Clouds. The system contains a WD and an M giant with an orbital period of 4.5 years (eg. Kato {\em et al.\/} \cite{kato}). It is a supersoft X-ray source powered by a steady hydrogen burning on the surface of the white dwarf (Orio {\em et al.\/} 2007). Because the system contains a massive white dwarf (M$_{WD}>$1.18 M$_\odot$; Orio {\em et al.\/} \cite{Orio}) with a high accretion rate ($\dot{\mathrm{M}}\simeq 10^{-7}$~M$_\odot$/yr; Kahabka \cite{kahabka}) it is considered as one of the most promising supernova Ia progenitors among the known symbiotic population.

\section{Model}
To estimate the mass of the red giant we used the fact that it pulsates with a period of 110 days (Kahabka \cite{kahabka}). SMC 3 lies on the sequence B in the K-log(P) plane (Wood 2000) which suggests the first overtone pulsation. Assuming the pulsation constant of Q=0.04 we derived the mass of 2.3$^{+0.6}_{-0.3}$~M$_\odot$.

To carry out our analysis we used the StarTrack population synthesis code (Belczy\'{n}ski {\em et al.\/} \cite{belczynski}). The code includes the wind accretion through  Bondi-Hoyle mechanism, Roche-lobe overflow,  $atmospheric$ Roche-lobe overflow and $wind$ Roche-lobe overflow. As initial conditions we adopted the current parameters of the system and then we modeled its future evolution. We assumed $\mathrm{M}_{\mathrm{Ch}}$=1.44~M$_\odot$ and, since we studied the system in the SN Ia context, a CO~WD.

\section{Results}
\begin{figure}[h]
\begin{center}
 \includegraphics[width=0.32\textwidth]{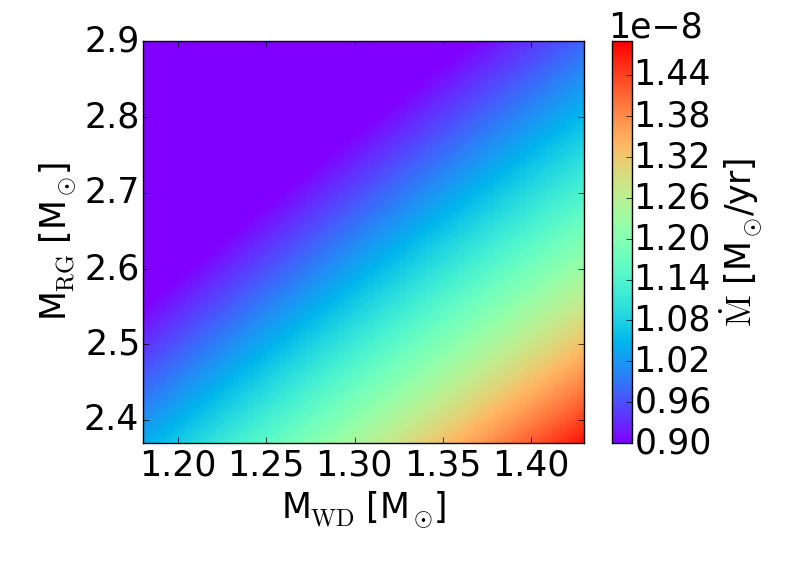}
 \includegraphics[width=0.32\textwidth]{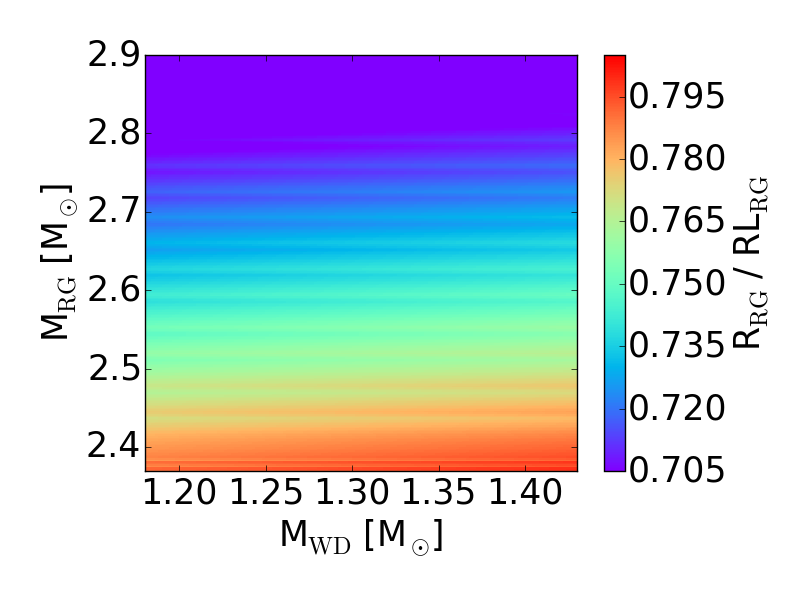}
 \includegraphics[width=0.32\textwidth]{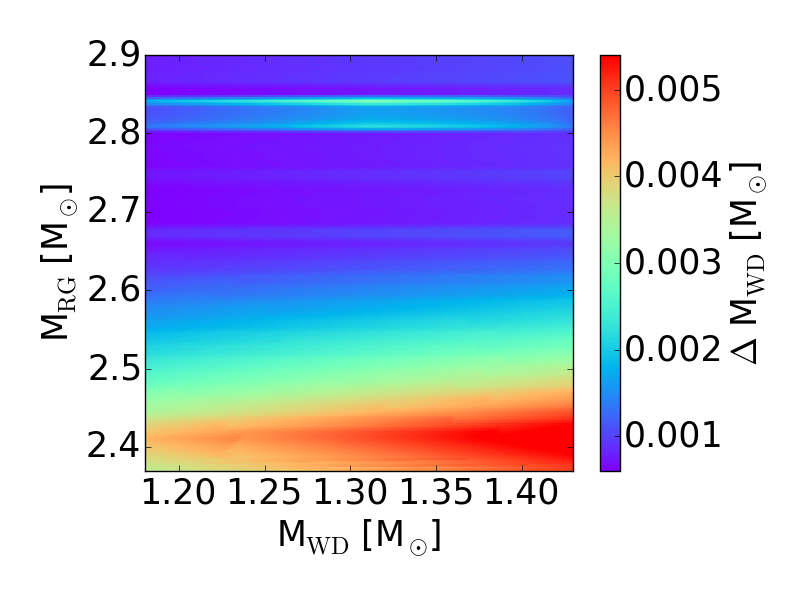}
\end{center}
\caption{Left panel: Predicted accretion rate on the WD. Middle panel: Roche lobe filling factor of the red giant. Right panel: Predicted mass growth of the WD before CE. }
\label{fig1}
\end{figure}

For all of our models the system went trough a common envelope (CE) after $\sim 10^5$~yrs and for none of the models system WD managed to accumulate enough matter to become a type Ia supernova. The orbital separation after the CE was relatively big (a$\simeq50-200$R$_\odot$), which makes a merger in the Hubble time unlikely. The obtained parameters of the system are presented in Fig.~\ref{fig1}.

\section{Conclusions}
We predict that the system will not become a type Ia supernova in contrary to what was suggested in the literature (Orio {\em et al.\/} \cite{Orio}). The caveat is that our predicted mass transfer rate is somewhat lower then the one expected for the steady hydrogen burning on the surface of the white dwarf (Nomoto {\em et al.\/} \cite{Nomoto}). A lower $\dot{\mathrm{M}}$ then expected could be due to the fact that our model underestimates the RG mass loss trough wind by treating it as in the single star scenario, whereas there is a strong observational evidence that this wind is significantly enhanced due to tidal interactions in SySt (Miko\l{}ajewska {\em et al.\/} \cite{mikolaj}).  Large Roche lobe filling factor suggests ellipsoidal variability in the system.




\begin{thebibliography}{99}
\bibitem[2008]{belczynski} Belczy\'{n}ski K. \etal\ 2008, ApJS, 174, 223
\bibitem[2004]{kahabka} Kahabka P., 2004, A\&A, 416, 57
\bibitem[2013]{kato} Kato M. \etal\ 2013, ApJ, 763, 5
\bibitem[2007]{Nomoto} Nomoto K. \etal\ 2007, ApJ, 663, 1269
\bibitem[2002]{mikolaj} Miko\l{}ajewska J. \etal\ 2002, AdSpR, 30, 2045
\bibitem[2007]{Orio} Orio M. \etal\ 2007, ApJ, 661, 1105
\bibitem[2000]{Wood} Wood P.R., 2000, PASA, 17, 18
\end{thebibliography}
\end{document}